\documentclass[aps,prl,twocolumn,amsmath,amssymb,superscriptaddress]{revtex4}
\usepackage{graphicx}
\usepackage{amssymb}
\usepackage{natbib}
\usepackage{color}

\newcommand{\beq}{\begin{eqnarray}}
\newcommand{\eeq}{\end{eqnarray}}

\begin{document}

\title{Coexistence of ferromagnetic and stripe antiferromagnetic spin fluctuations
in SrCo$_2$As$_2$}
\author{Yu Li}
\affiliation{Department of Physics and Astronomy,
Rice University, Houston, Texas 77005, USA}
\affiliation{Department of Physics, Beijing Normal University, Beijing 100875, China}

\author{Zhiping Yin}
\email{yinzhiping@bnu.edu.cn}
\affiliation{Department of Physics, Beijing Normal University, Beijing 100875, China}

\author{Zhonghao Liu}
\email{lzh17@mail.sim.ac.cn}
\affiliation{State Key Laboratory of Functional Materials for Informatics and Center for Excellence in Superconducting Electronics, SIMIT, Chinese Academy of Sciences, Shanghai 200050, China}

\author{Weiyi Wang}
\affiliation{Department of Physics and Astronomy,
Rice University, Houston, Texas 77005, USA}

\author{Zhuang Xu}
\affiliation{Department of Physics, Beijing Normal University, Beijing 100875, China}

\author{Yu Song}
\affiliation{Department of Physics and Astronomy,
Rice University, Houston, Texas 77005, USA}

\author{Long Tian}
\affiliation{Department of Physics, Beijing Normal University, Beijing 100875, China}

\author{Yaobo Huang}
\affiliation{Shanghai Synchrotron Radiation Facility, Shanghai Institute of Applied Physics, Chinese Academy of Sciences, Shanghai 201204, China}

\author{Dawei Shen}
\affiliation{State Key Laboratory of Functional Materials for Informatics and Center for Excellence in Superconducting Electronics, SIMIT, Chinese Academy of Sciences, Shanghai 200050, China}

\author{D. L. Abernathy}
\affiliation{Neutron Scattering Division, Oak Ridge National Laboratory, Oak Ridge, Tennessee 37831, USA}

\author{J. L. Niedziela}
\affiliation{Neutron Scattering Division, Oak Ridge National Laboratory, Oak Ridge, Tennessee 37831, USA}

\author{R. A. Ewings}
\affiliation{ISIS Pulsed Neutron and Muon Source, STFC Rutherford Appleton Laboratory, Didcot, Oxfordshire, OX11 0QX, UK}

\author{T. G. Perring}
\affiliation{ISIS Pulsed Neutron and Muon Source, STFC Rutherford Appleton Laboratory, Didcot, Oxfordshire, OX11 0QX, UK}

\author{Daniel Pajerowski}
\affiliation{Neutron Scattering Division, Oak Ridge National Laboratory, Oak Ridge, Tennessee 37831, USA}

\author{Masaaki Matsuda}
\affiliation{Neutron Scattering Division, Oak Ridge National Laboratory, Oak Ridge, Tennessee 37831, USA}

\author{Philippe Bourges}
\affiliation{Laboratoire L$\acute{e}$on Brillouin, CEA-CNRS, Universit$\acute{e}$ Paris-Saclay, CEA Saclay, 91191 Gif-sur-Yvette, France}

\author{Enderle Mechthild}
\affiliation{Institut Laue-Langevin, 6 rue Jules Horowitz, Bo$\hat{i}$te Postale 156, 38042 Grenoble Cedex 9, France}

\author{Yixi Su}
\affiliation{J$\ddot{u}$lich Centre for Neutron Science, Forschungszentrum Jülich GmbH, Outstation at MLZ, D-85747 Garching, Germany}

\author{Pengcheng Dai}
\email{pdai@rice.edu}
\affiliation{Department of Physics and Astronomy,
Rice University, Houston, Texas 77005, USA}
\affiliation{Department of Physics, Beijing Normal University, Beijing 100875, China}
\date{\today}

\begin{abstract}
We use inelastic neutron scattering to study energy and wave vector dependence of spin fluctuations in SrCo$_2$As$_2$, derived from
SrFe$_{2-x}$Co$_x$As$_2$ iron pnictide superconductors. Our data reveals
the coexistence of antiferromagnetic (AF) and ferromagnetic (FM) spin fluctuations at wave vectors $\textbf{Q}_{\rm AF}$=(1,0) and $\textbf{Q}_{\rm FM}$=(0,0)/(2,0), respectively.
By comparing neutron scattering results with those of dynamic mean field theory calculation and angle-resolved photoemission spectroscopy experiments, we conclude that both AF and FM spin fluctuations
in SrCo$_2$As$_2$
 are closely associated with a flat band of the $e_g$ orbitals near the Fermi level, different
from the $t_{2g}$ orbitals in superconducting SrFe$_{2-x}$Co$_x$As$_2$.
Therefore, Co-substitution in SrFe$_{2-x}$Co$_x$As$_2$
induces a $t_{2g}$ to $e_g$ orbital switching, and is responsible for
FM spin fluctuations detrimental to the singlet pairing superconductivity.
\end{abstract}

\maketitle

Flat electronic bands can give rise to a plethora of interaction-driven quantum phases, including
ferromagnetism \cite{tasaki}, Mott insulating phase due to electron correlations \cite{Cao1},
and superconductivity \cite{Cao2}. Therefore, an understanding how the flat electronic bands can influence the electronic, magnetic, and superconducting properties of solids is an important topic in condensed matter physics.  In iron pnictide superconductors such as
$A$Fe$_{2-x}$Co$_x$As$_2$ ($A=$ Ba, Sr) [Figs. 1(a)-1(d)], the dominate interactions are stripe antiferromagnetic
(AF) order, and superconductivity, which has singlet electron pairing, arises by doping electron with Co-substitution to suppress static AF order \cite{Johnston,Scalapino2012,RMP_Dai}.
While AF spin fluctuations and superconductivity in iron pnictides are believed to arise from nested hole Fermi surfaces
at $\Gamma$ and electron Fermi surfaces at $M$ [Fig. 1(e)] \cite{Hirschfeld11}, the
density functional theory (DFT) calculations suggest the competing
ferromagnetic (FM) and AF spin fluctuations with the balance controlled by
doping \cite{DFT_fm1,DFT_fm2}.
For Co-overdoped $A$Co$_2$As$_2$ \cite{Sefat09,Pandey13}, where the
DFT calculations find a tendency for both the FM
and AF order, neutron scattering revealed only the AF
spin fluctuations \cite{SCA_stripe} while angle resolved photoemission spectroscopy (ARPES) experiments found no evidence of the Fermi surface nesting \cite{NXu13,Dhaka13}. On the other hand, nuclear magnetic resonance (NMR) measurements on $A$Fe$_{2-x}$Co$_x$As$_2$ provided evidence
for FM spin fluctuations at all Co-doping levels in addition to the
AF spin fluctuations \cite{NMR_SCA,NMR_FM}.
In particular, strong FM spin fluctuations in
$A$Fe$_{2-x}$Co$_x$As$_2$ are believed to compete with AF spin fluctuations and prevent superconductivity for Co-overdoped samples \cite{NMR_SCA,NMR_FM}, contrary to the Fermi
surface nesting picture where superconductivity is suppressed via vanishing hole Fermi surfaces with
increasing Co-doping \cite{Hirschfeld11,Meng2013}. Finally, action of physical, chemical pressure,
or aliovalent substitution in $B$Co$_2$As$_2$ ($B=$ Eu, Ca) can drive these
AF materials into ferromagnets \cite{XTan16}. In particular,
CaCo$_{1.84}$As$_2$ with a collapsed tetragonal structure \cite{Kreyssig08} forms
 A-type AF ground state with coexisting FM spin fluctuations within the CoAs layer and A-type AF spin fluctuations between the CoAs layers \cite{Sapkota2017}.  These features are different from those of Ca(Fe$_{1-x}$Co$_{x}$)$_2$As$_2$ \cite{Sapkota18,jzhao09} and $A$Fe$_{2-x}$Co$_x$As$_2$ \cite{RMP_Dai}.

\begin{figure}[htb]
\includegraphics[scale=.33]{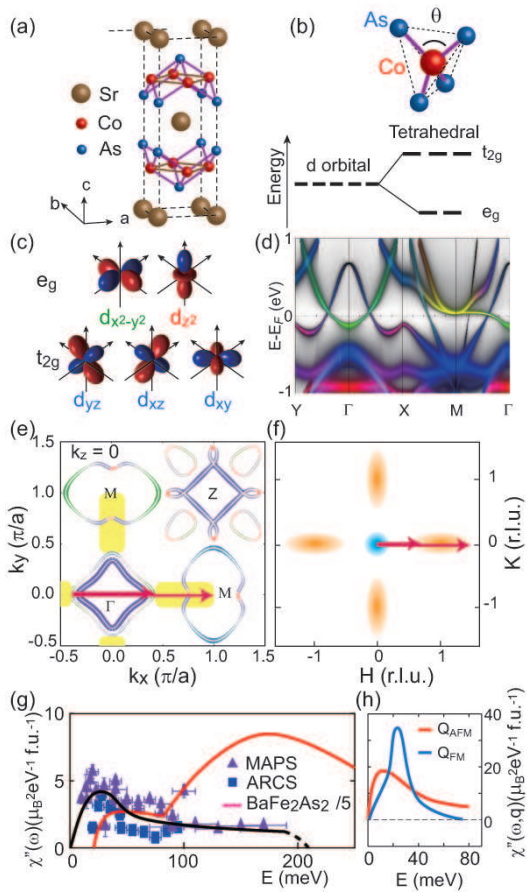}
\caption{(a) Crystal structure of SrCo$_2$As$_2$. (b) The tetrahedron of Fe(Co)As$_4$ and the resulting $d$-orbital splitting. (c) Wave functions of the five $d$-orbitals. (d) Band structure of SrCo$_2$As$_2$. Green(Red) represents $d_{x^2-y^2}$($d_{z^2}$) orbital and blue is the contribution from the $t_{2g}$ ($d_{xz}$, $d_{yz}$, $d_{xy}$) orbitals. Yellow is the mixture of red ($d_{z^2}$) and green ($d_{x^2-y^2}$). (e) Fermi surfaces from DFT+DMFT calculations. The shading yellow area corresponds to the flat band (yellow part) in Fig. 1(d) and arrows represent scattering wave vectors associated with the flat band. The colors represent the same orbital characters as in (c) and (d). (f) Schematics of the low energy FM (blue) and AF (orange) spin fluctuations in SrCo$_2$As$_2$. (g) Energy dependence of integrated $\chi^{\prime\prime}(E)$ of SrCo$_2$As$_2$ in absolute units normalized by using a vanadium 
standard \cite{SI}. The red solid line is $\chi^{\prime\prime}(E)/5$ of BaFe$_2$As$_2$ \cite{leland11}. The black solid line is a guide to the eye. (h) The measured AF and FM fluctuations at ${\bf Q}_{\rm AF}$ and ${\bf Q}_{\rm FM}$ \cite{SI}.
}
\end{figure}

Iron pnictides have five nearly degenerate $d$ orbitals
which split into $t_{2g}$ and $e_g$ orbitals in a tetrahedral crystal
field [Figs. 1(b), 1(c)].
The electronic structure of the system is dominated by Fe 3$d$ $t_{2g}$ orbitals near the Fermi level with hole-electron Fermi surfaces at $\Gamma$ and $M$, respectively [Fig. 1(e)]. The presence of multiple Fe 3$d$ orbitals near the Fermi level results in varying orbital characters on different parts of the Fermi surfaces \cite{MYi_npj_QM}, and orbital-dependent strengths of electronic correlations \cite{Yin2011,Medici2014,Nica17,QMSi_NRM,Zhonghao2018}. The electronic band structures of SrCo$_2$As$_2$ calculated
by the DFT combined with dynamic
mean field theory (DMFT) \cite{Kotliar06,Haule10} reveal the presence of a flat band near $M$ point with
mixture of the $d_{z^2}$ and $d_{x^2-y^2}$ orbitals [Fig. 1(d)]. If SrCo$_2$As$_2$ has strong ferromagnetism arising from the flat band as suggested from
NMR \cite{NMR_SCA,NMR_FM}, one should be able to extract its energy and wave vector
dependence by neutron scattering and determine its role to the suppressed superconductivity in Co-overdoped SrFe$_{2-x}$Co$_x$As$_2$ \cite{Johnston,Scalapino2012,RMP_Dai}.

In this Letter, we combine neutron scattering, ARPES and DFT+DMFT methods to study SrCo$_2$As$_2$, an electron-doped end member of SrFe$_{2-x}$Co$_x$As$_2$ exhibiting no structural, magnetic, or superconducting transitions \cite{Pandey13}.
Besides confirming the longitudinally elongated AF spin fluctuations at wave vector $\textbf{Q}_{\rm AF}$=(1,0) [Figs. 1(f) and 2] \cite{SCA_stripe}, we successfully observed the in-plane FM spin fluctuations at $\textbf{Q}_{\rm FM}$=(0,0) and its equivalent $(2,0)$ positions [Figs. 2 and 3]. From the DFT+DMFT calculations and ARPES measurements, we
find a flat band  consisting of the $e_g$ orbitals along the $\Gamma$-$M$ direction right above the Fermi level [Fig. 1(d)], leading to a prominent peak
in the density-of-state (DOS) near Fermi level responsible for both the FM and
AF spin fluctuations [Figs.  4(a)-4(d)]. Orbital analysis of the dynamic spin susceptibility $\chi^{\prime\prime}({\bf Q},E)$ in the DFT+DMFT calculations suggests that magnetism in SrCo$_2$As$_2$ is dominated by the $e_g$ orbitals [Figs. 1(d), 1(e), 4(e), 4(f)]. These results are beyond the prevailing orbital selective Mott picture in iron pnictides,
where the $t_{2g}$ orbitals are most strongly correlated \cite{MYi_npj_QM,QMSi_NRM,Chenglin_NaFeAs,YL_LiFeAs,YSong16} and electron (Co) doping monotonously reduces correlations in all five $d$ orbitals \cite{Yin2011,Medici2014}. In addition, the FM spin correlations in SrCo$_2$As$_2$ are similar to the A-type AF order 
in CaCo$_{1.86}$As$_2$ \cite{AtypeCaCo2As2}. Therefore, our observation is consistent with the proposal that FM fluctuations are detrimental to superconductivity in Co-overdoped $A$Fe$_{2-x}$Co$_x$As$_2$ and may be responsible for the hole-electron asymmetry of the superconducting dome in iron pnictide families \cite{NMR_FM}.

\begin{figure}[t]
\includegraphics[scale=.36]{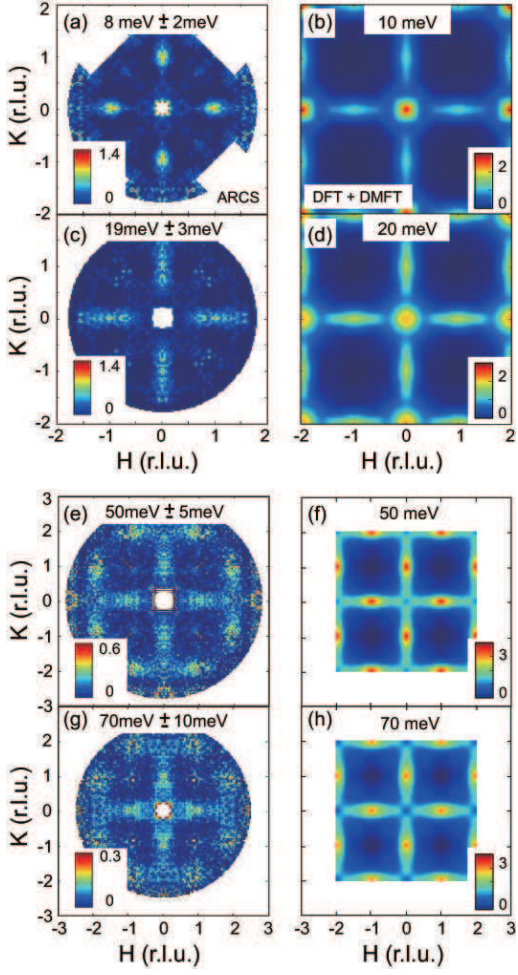}
\caption{(a,c,e,g) Two-dimensional images of measured dynamic spin susceptibility of SrCo$_2$As$_2$ in the $[H,K]$ plane at $E=8\pm 2$, $19\pm 3$, $50\pm 5$, and $70\pm 10$ meV, respectively. Radially symmetric backgrounds were subtracted to visually enhance the weak magnetic signal. (b,d,f,h) The corresponding results from the DFT+DMFT calculations \cite{SI}.
}
\end{figure}

We begin by showing constant-energy slices of $\chi^{\prime\prime}({\bf Q},E)$ 
on SrCo$_2$As$_2$ at $T = 5$ K [Figs. 2(a),(c),(e),(g)] \cite{SI,YL_LiFeAs}. At $E = 8$ meV, the AF spin fluctuations at ${\bf Q}_{\rm AF} = (1,0)$ are longitudinally elongated similar to that in hole-doped BaFe$_2$As$_2$ [Fig. 2(a)] \cite{Meng2013}. With increasing energy, spin fluctuations along the longitudinal direction are further elongated  while they barely change along the transverse direction, different from the transversely 
elongated spin fluctuations in $A$Fe$_{2-x}$Co$_x$As$_2$ \cite{RMP_Dai,Meng2013}.
 At $E\geq 50$ meV, there are magnetic
intensities at both the $\textbf{Q}_{\rm AF} = (1,0)$ and $\textbf{Q}_{\rm FM} = (2,0)$.
Spin fluctuations form ridges of scattering across the whole Brillouin zone (BZ) forming a square network [Figs. 2(e), 2(g)], similar to those in CaCo$_{2-y}$As$_2$ \cite{Sapkota2017}.
Along the transverse direction, we observed a linearly broadening of the half-width at half-maximum (HWHM) of AF spin fluctuations with increasing energy at the speed of $\Delta HWHM / \Delta E \approx 1/(440$ meV$\cdot$ {\AA}) \cite{SI} and no peak splitting was identified.

We used the DFT+DMFT calculations to understand the electronic band structure [Fig. 1(d)]
and spin dynamics of SrCo$_2$As$_2$ \cite{Yin2011,SI,ZPYin14}. Figures 2(b), 2(d), 2(f) and 2(h)
show the DFT+DMFT calculated results for $E=10, 20, 50, 70$ meV.
Although the calculated results look remarkably similar to experimental data in Figs. 2(a), 2(c), 2(e), and 2(g),
there are also important differences. First, the AF spin fluctuations are weaker than the FM spin fluctuations in the DFT+DMFT calculation at $E=10$ meV, while they are stronger in experiments. This is mostly because the calculations are exceedingly sensitive to the position of the flat band with respect to the Fermi level.  Second,
the calculation suggests that FM spin fluctuations originating from $\Gamma$ (and equivalent) point
merge into AF spin fluctuations at ${\bf Q}_{\rm AF}=(\pm1,0)/(0,\pm 1)$ around 50 meV [Fig. 2(f)],
while there is no clear evidence of FM spin fluctuations at $E=8,19$ meV [Figs. 2(a), 2(c)] \cite{SI}. Figure 1(g) shows energy dependence of local dynamic susceptibility
$\chi^{\prime\prime}(E)$, obtained by integrating both the FM and AF signal
within the area of $(0,0)\rightarrow (1,1)\rightarrow (2,0)\rightarrow (1,-1)\rightarrow (0,0)$ \cite{RMP_Dai}, and
its comparison with those of BaFe$_2$As$_2$ \cite{leland11}. The total fluctuating moment is approximately $\langle m^2 \rangle \approx 0.4 \pm 0.1\ \mu^2_B$/f.u. \cite{leland11,SI}, compared with 0.5 $\mu^2_B$/f.u. from the calculation. Due to the diffusive nature of the magnetic scattering, it is rather difficult to experimentally separate the integrated FM and AF signal and compare with that of the DFT+DMFT calculations. 

\begin{figure}[t]
\includegraphics[scale=.32]{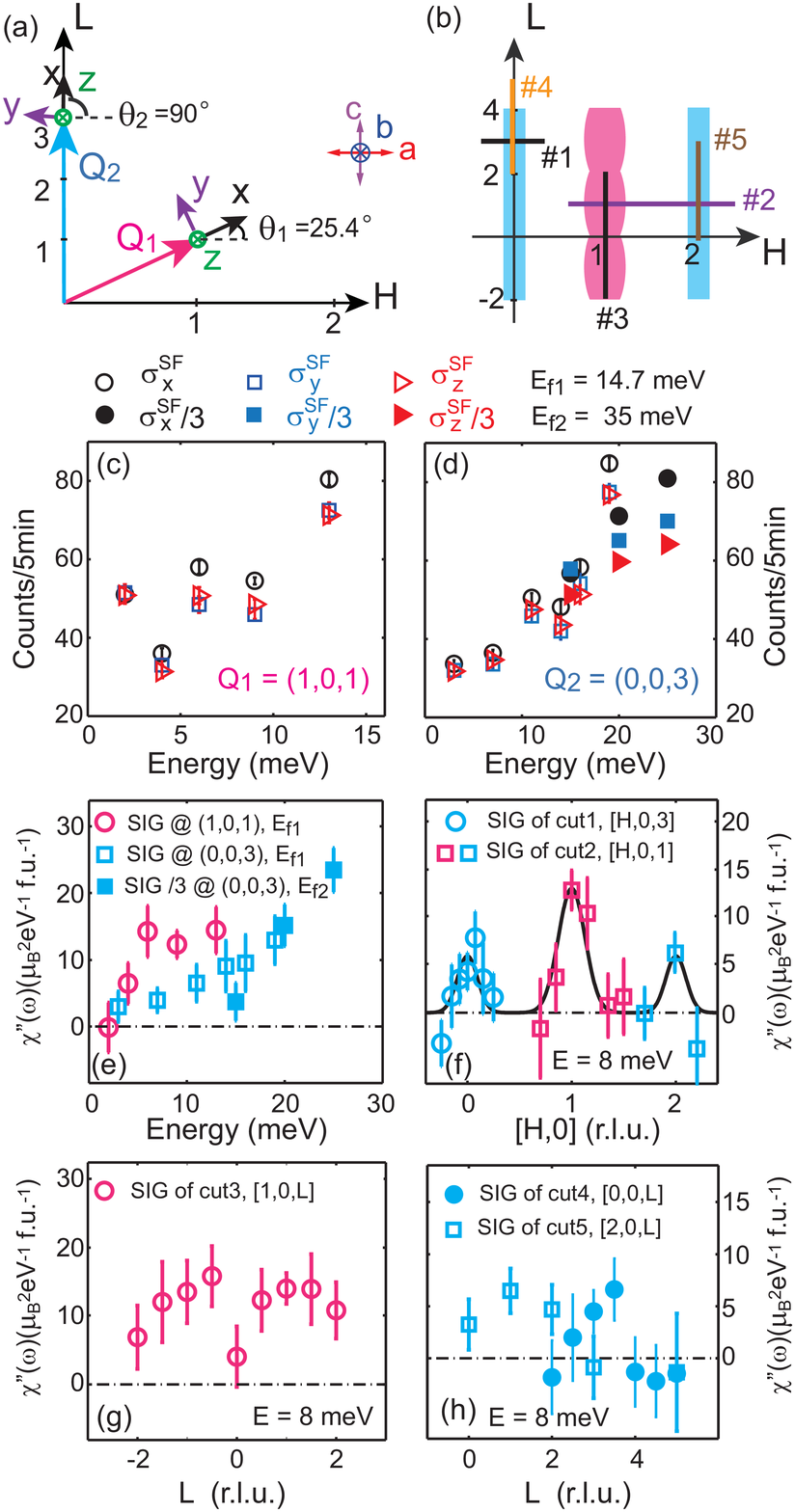}
\caption{(a) Schematic of the $[H,0,L]$ scattering plane for neutron polarization analysis.
The AF and FM wave vectors are labeled as ${\bf Q}_1$ and ${\bf Q}_2$, respectively.
The neutron polarization directions are along the $x$, $y$, and $z$.
(b) Locations of FM (blue) and AF (magenta) spin fluctuations in reciprocal space.
Lines indicate scan directions.
(c,d) Constant-${\bf Q}$ scans of $\sigma_x^{SF}$, $\sigma_y^{SF}$,
and $\sigma_z^{SF}$ at ${\bf Q_1}$ and ${\bf Q_2}$, respectively,
at $T = 1.5$ K. (e) Constant-$Q$ scans of pure magnetic scattering intensity
at ${\bf Q_1}$ and ${\bf Q_2}$. (f,g,h) The AF (magenta) and FM (blue) scattering 
at $E = 8$ meV along the $H$ and $L$ directions as marked in (b).  The values of $SIG$ are converted into absolute units by assuming the polarized data at ${\bf Q}_{\rm AF}= (1,0,1)$ and $E=8$ meV  is comparable with the integrated intensity in $0.975<$H$<1.025$ and $-0.1<$K$<0.1$ in Fig. 2(a).
}
\end{figure}

To conclusively determine the FM signal in SrCo$_2$As$_2$,
we carried out polarized neutron scattering experiments with
the neutron polarization directions $x$, $y$, and $z$ shown in Fig. 3(a), which 
correspond to neutron spin-flip (SF) scattering cross sections
$\sigma_x^{SF}$, $\sigma_y^{SF}$, and $\sigma_z^{SF}$, respectively \cite{moon,polarized1,polarized2,polarized3,polarized4,YL_BaFe2As2}. 
The magnetic scattering of SrCo$_2$As$_2$ should then be $SIG=\sigma_x^{SF}- (\sigma_y^{SF}+ \sigma_z^{SF})/2$ \cite{polarized1,polarized2,polarized3,polarized4,YL_BaFe2As2}.
Figures 3(c) and 3(d) show the energy scans 
at $\textbf{Q}_1 = (1,0,1)$ and $\textbf{Q}_2 = (0,0,3)$ [Fig. 3(a)].
Figure 3(e) shows energy
dependence of $SIG$ at $\textbf{Q}_1$ and $\textbf{Q}_2$, confirming the presence of magnetic fluctuations
at the AF and FM wave vectors, respectively.

At $\textbf{Q}_1$ [Fig. 3(c)],
 $\sigma_y^{SF} \approx \sigma_z^{SF}$ implies that the AF spin fluctuations are isotropic in spin space, different from the anisotropic spin fluctuations in
BaFe$_{2-x}$Co$_x$As$_2$ induced by spin-orbit coupling \cite{polarized1,polarized2,polarized3,polarized4,YL_BaFe2As2}. These results suggest that
the spin-orbit coupling in SrCo$_2$As$_2$ is weaker than that of BaFe$_2$As$_2$. 
At $\textbf{Q}_2$ [Figs. 3(d), 3(e)], magnetic scattering increases 
with increasing energy with no spin gap above $E=3$ meV, providing direct evidence
for the FM spin fluctuations in SrCo$_2$As$_2$ \cite{NMR_SCA,SI}. To further demonstrate the coexisting FM
and AF spin fluctuations, we performed constant-energy 
scans along the $[H,0,3]$ and $[H,0,1]$
directions at $E=8$ meV [Fig. 3(b)]. Figure 3(f) indicates that the FM spin fluctuations are confined
near $(0,0,3)$ and are about half the size as that of the
AF signal around $(1,0,1)$.  The DFT+DMFT calculations
predict the dominant FM spin fluctuations
around 10 meV [Fig. 2(b)]. Constant-energy scans along the
$[1,0,L]$ [Fig. 3(g)] and $[0,0,L]$ [Fig. 3(h)] directions reveal weakly $L$ dependent scattering
at both the AF and FM positions, respectively, confirming the quasi-two-dimensional nature of
the magnetic scattering. Figure 1(h) shows energy dependence of $\chi^{\prime\prime}({\bf Q},E)$ at ${\bf Q}_{\rm AF}$ and ${\bf Q}_{\rm FM}$, 
where the peak in ${\bf Q}_{\rm FM}$ near 25 meV should be associated with the Van Hove singularity of the flat band. 

\begin{figure}[t]
\includegraphics[scale=.25]{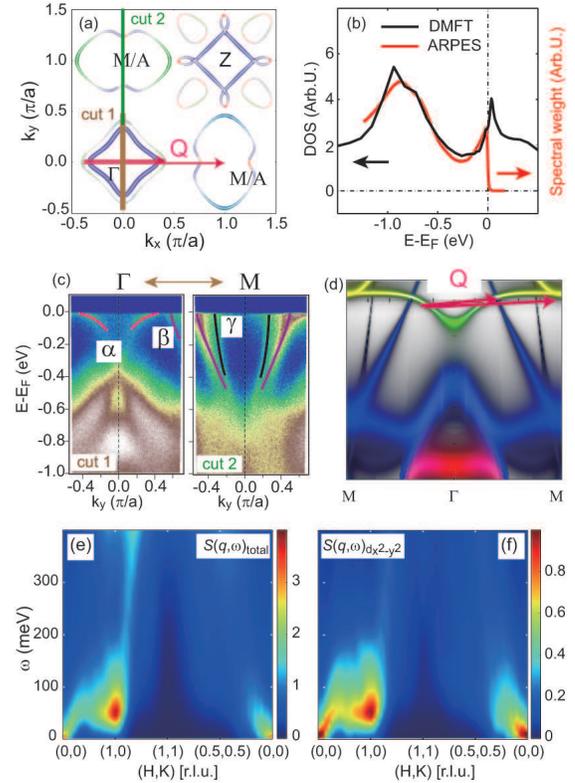}
\caption{(a) Fermi surfaces of SrCo$_2$As$_2$ from the DFT+DMFT calculations. (b) Calculated electronic DOS and integrated spectral weight from ARPES. (c) Intensity plots of the band dispersion
along the $\Gamma$-$M$ direction ($E_{photon} = 22$ eV). (d) Calculated band structure
along the $M$-$\Gamma$-$M$ direction. Arrows indicate possible
wave vectors from occupied to empty states on the flat band. (e) Total $\chi^{\prime\prime}({\bf Q},E)$ from the DFT+DMFT calculation. (f) Calculated $\chi^{\prime\prime}({\bf Q},E)$ from the $d_{x^2-y^2}$ orbital.
}
\end{figure}

To understand the origin of the FM and AF spin fluctuations in SrCo$_2$As$_2$ [Fig. 4(a)], we measured its band structure by ARPES and compared the outcome in Fig. 4(c) with the DFT+DMFT calculations in Fig. 4(d). Around the $\Gamma$ point, one shallow electron-like $\alpha$ band
and one highly dispersive hole-like $\beta$ band were observed.
Another electron-like band at the $M$ point was also found.
These results agree well with the DFT+DMFT calculation in Fig. 4(d), supporting the existence of
a flat band along $\Gamma$-$M$ direction right above the Fermi level [Figs. 1(d) and 4(d)] \cite{SI}. 
Further ARPES data collected along
the $Z$-$A$ direction with a different photon energy reveals the presence of the flat band (or band bottom) touching the Fermi level at $A$ point, mainly arising from the $d_{x^2-y^2}$
orbital hybridized with the $d_{z^2}$ [Fig. 1(d)] \cite{SI}. In particular, the partial DOS of the Co 3$d_{x^2-y^2}$ orbital in the DFT+DMFT calculation exhibits a peak at about 35 meV above the Fermi level, similar to the maximum scattering of the FM spin fluctuations [Fig. 1(h)], suggesting a close relationship between the flat band and FM instability. 

Flat electronic bands with high DOS near the Fermi level can
 influence the electronic and magnetic properties of solids through tuning the
electron-electron correlations \cite{tasaki,Cao1,Cao2}.
In SrCo$_2$As$_2$, the flat band might affect spin fluctuations in two ways.
First, the $d_{x^2-y^2}$ band ($\alpha$) dispersive along the $\Gamma$-$X/Y$ direction
but flat along the $\Gamma$-$M$ direction [Fig. 1(d)] might lead to high DOS
near the Fermi level and Stoner FM instability similar to that of
 Sr$_2$RuO$_4$ \cite{RMP214,FM214}.  Both the DFT+DMFT calculations and ARPES experiments reveal a prominent
peak in DOS near the Fermi level [Fig. 4(b)], supporting the existence of flat-band related FM  fluctuations. Second, the flat band above the Fermi level provides
many electron scattering channels as shown by the arrows in Fig. 4(d). These scattering processes
result in the longitudinally elongated spin fluctuations extending from $\Gamma$ to $M$ [Fig. 1(f)].  This
is different from the longitudinally elongated low-energy spin fluctuations in hole-doped BaFe$_2$As$_2$, where
the longitudinal elongation is driven by mismatched sizes of the hole-electron
Fermi surfaces \cite{Meng2013,JHZhang10,CLZhang11,RZhang18}.
Figures 4(e) and 4(f) plot the DFT+DMFT calculated total dynamic spin susceptibility
and contributions from the $d_{x^2-y^2}$ orbital \cite{SI}. Surprisingly, both the AF and FM spin
fluctuations are dominated by the $e_g$ orbitals (Fig. S5) \cite{SI}, different from the majority $t_{2g}$ contributions to the spin dynamics in iron pnictides \cite{ZPYin14}. In  SrFe$_{2-x}$Co$_x$As$_2$, the presence of AF spin fluctuations \cite{SCA_stripe} is responsible for the superconductivity. The appearance of FM spin fluctuations in SrCo$_2$As$_2$ and their competition with the stripe AF spin fluctuations might be responsible the absence of superconductivity in heavily over-doped SrFe$_{2-x}$Co$_x$As$_2$. The underlying orbital characters might also be an important factor for superconductivity in iron pnictides.

The work at Rice is supported by the US NSF DMR-1700081 and the Robert
A. Welch Foundation grant No. C-1839 (P.D.). ZPY was supported by the NSFC (Grant No. 11674030),  the Fundamental Research Funds for the Central Universities (Grant No. 310421113),  the National Key Research and Development Program of China grant 2016YFA0302300. The calculations used high performance computing clusters at BNU in Zhuhai and the National Supercomputer Center in Guangzhou. ZHL acknowledges the NSFC (Grant No. 11704394), and the Shanghai Sailing Program (Grant No.17YF1422900).
We acknowledge the support of the HFIR/SNS, a DOE User Facility operated by ORNL. Experiments at the ISIS were supported by a beam time allocation from the STFC.

{}

\end{document}